# Growth of [001]-oriented polycrystalline Heusler alloy thin films using [001]-textured Ag buffer layer on thermally oxidized Si substrate for spintronics applications

Dolly Taparia, Taisuke T. Sasaki,[*] Tomoya Nakatani,[**] Hirofumi Suto, Seiji Mitani, and Yuya Sakuraba

Research Center for Magnetic and Spintronic Materials, National Institute for Materials Science, Tsukuba, 305-0047, Japan.

**Abstract**

To utilize half-metallic Heusler alloys in practical spintronic devices, such as magnetic sensors and magnetic memories, the key is to realize highly textured and structurally ordered polycrystalline thin films. In this study, we fabricated polycrystalline $Co_2FeGa_{0.5}Ge_{0.5}$ (CFGG) Heusler alloy films deposited on a [001]-oriented Ag buffer layer, which was achieved by introducing $N_2$ into Ar during the sputtering process, on a thermally oxidized Si substrate. We obtained strongly [001]-oriented CFGG films with B2 ordering and a high saturation magnetization close to the theoretical value, which can provide highly spin-polarized electric and spin current sources in spintronic devices with industrial viability.

[*],[**]Corresponding authors: SASAKI.Taisuke@nims.go.jp, NAKATANI.Tomoya@nims.go.jp

# I. Introduction

The field of spintronics leverages the interplay between the two fundamental properties of electron – *spin* and *charge*, leading to various technological advancements, from magnetic sensors to reconfigurable magnetic logic circuits.[1, 2] Highly spin-polarized currents are crucial for achieving higher magnetoresistance (MR) ratios in tunnel magnetoresistance and current-perpendicular-to-plane giant magnetoresistance (CPP-GMR) devices, which are important for the application of e.g., magnetic sensors and spin-transfer-torque magnetic random-access memory. For example, the enhancement of the MR ratio of CPP-GMR devices is essential for their application to the read heads of hard disk drives with ultra-high recording densities. This has driven extensive research efforts focused on finding materials with high spin polarization, ultimately half-metals with 100% spin polarization.[3, 4]

Co-based Heusler alloys with a chemical formula of $Co_2YZ$ (e.g., $Y$ = Mn and Fe, and $Z$ = Al, Si, Ga, and Ge) have received intensive research attention as a promising candidate for half-metal operating at room temperature (RT) due to their high Curie temperature.[5] Using fully epitaxial $Co_2MnSi$, $Co_2Fe_{0.4}Mn_{0.6}Si$, and $Co_2FeGa_{0.5}Ge_{0.5}$ Heusler alloy films deposited on a MgO (001) single-crystalline substrate via an appropriate buffer layer, large CPP-GMR ratios up to 82% have been achieved.[6-8] Because the usage of single-crystalline substrates limits the industrial viability of the devices, achieving high spin polarization in polycrystalline Heusler alloy thin films is critical. Since the body-centered-cubic (bcc)-based Heusler alloys have the natural crystallographic texture along the [110] orientation, many reports have focused on the [110]-oriented polycrystalline Heusler alloy thin films for the CPP-GMR device applications. A maximum MR of 25% has been reported in the CPP-GMR devices with [110]-oriented polycrystalline $Co_2Mn_{0.6}Fe_{0.4}Ge$ films based CPP-GMR.[9]

[111]-oriented spacer layers with the face-centered-cubic (fcc) structure, e.g., Cu, Ag, and AgSn, can be grown on the [110]-oriented Heusler alloy films with the orientation relation of $(011)[100]_{Heusler} \parallel (111)[1\bar{1}0]_{fcc}$.[10] While the lattice misfit of the interface along the $\langle 100 \rangle_{Heusler}$ and $\langle 110 \rangle_{fcc}$ is small, that along the $\langle 110 \rangle_{Heusler}$ and $\langle 112 \rangle_{fcc}$ is much larger, e.g., 23.3% for $Co_2FeGa_{0.5}Ge_{0.5}$ (CFGG, $a$ = 0.4086 nm [11]) and Ag. Such a large lattice misfit at the interface may give rise to spin-independent scattering of the conducting electron, which results in a reduction in MR ratio. One the other hand, the interface between [001]-textured Heusler alloy and [001]-textured fcc spacer maintains a small lattice misfit with the orientation relationship of $(001)[110]_{Heusler} \parallel (111)[100]_{fcc}$. Therefore, the purpose of the present work is to obtain [001]-oriented polycrystalline CPP-GMR structure. Although a [001]-textured MgO buffer layer was previously employed to achieve [001]-oriented polycrystalline CPP-GMR devices, the insulating nature of the MgO buffer renders it unsuitable for the fabrication of actual read sensors, which require a conductive underlayers.[12]

Ag is a suitable buffer layer material for CPP-GMR devices due to its low electrical resistivity, good thermal conductivity, low diffusivity into other metals, and good lattice matching with most Heusler alloys.[13-15] For many fcc metals, the addition of $N_2$ and $O_2$ in Ar for sputtering deposition was reported to change the out-of-plane crystallographic orientation from [111] to [$\bar{0}$01]-direction.[16-20] The drawback of this method is the incorporation of $O_2$ or $N_2$ in the fcc films, as found in the Pt growth in Ar+$O_2$ [16] and the Ni growth in Ar+$N_2$.[19] In the case of Ag, Hu *et al.* demonstrated the growth of highly [001]-orientated Ag films on (001) single crystal Si sheets with thicknesses ranging from 1.6 to 1.9 μm in an Ar+$N_2$ atmosphere. [20] They studied the $N_2$ concentration dependence of the texture of the Ag film and found [001]-oriented Ag films without any nitride phase. However, scanning electron microscopy (SEM) imaging revealed holes and big spherical crystalline granules on the film surface of such µm-level thick [001]-oriented Ag films, which is not suitable for a buffer layer for spintronic applications. Therefore, there have been no reports on utilizing such [001]-oriented Ag films as a buffer layer for the growth of [001]-oriented polycrystalline ferromagnetic layers for spintronic devices, such as CPP-GMR. In this study, we fabricated a flat [001]-oriented Ag buffer layer using Ar+$N_2$ sputtering gas and successfully fabricated a [001]-oriented CFGG Heusler alloy thin film on a thermally oxidized Si substrate, which is suitable for industrial applications.

## II. Experimental details

Ag thin films of 100 nm thickness were deposited on a thermally oxidized Si (Si/$SiO_2$) substrate by RF magnetron sputtering in an Ar+$N_2$ atmosphere. To enhance the adhesion of the Ag film with the substrate and protect against surface oxidization, a 5 nm Ta seed layer and a 2 nm Ta capping layer were deposited (Fig. 1(a)). The base pressure of the sputtering chamber was less than $1\times10^{-7}$ Pa. The working pressure for the deposition of Ag was maintained at 7 mTorr. The Ar flow rate was kept constant at 10 sccm, and the $N_2$ flow rate was varied to 0, 5, 15, 20 and 25 sccm. Subsequently, a film stack of Si/$SiO_2$/Ta (5 nm)/Ag (100 nm)/CFGG (20 nm)/Ta (2 nm) was prepared to study the growth of the CFGG Heusler thin film on the Ag buffer layer. All the films were deposited at room temperature and *ex-situ* annealed at $T_A$ = 300 °C for 30 min under high vacuum condition. The crystal structures of the samples were examined using X-ray diffraction (XRD) with the Cu-$K_\alpha$ radiation ($\lambda$ = 1.5406 Å). The composition of the CFGG film was measured to be $Co_{47.6}Fe_{24.8}Ga_{15.0}Ge_{12.6}$ (in atomic %) by X-ray fluorescence spectroscopy. A comprehensive microstructural analysis of the CFGG layer and the Ag buffer layer was performed using a spherical aberration-corrected (scanning) transmission electron microscope ((S)TEM) using FEI Titan G2 80-200). The specimen preparation for the (S)TEM observation was carried out *via* the lift-out technique employing a dual-beam focused ion beam/scanning electron microscope (FIB/SEM, FEI Helios G4). The magnetic properties of the CFGG Heusler alloy thin films were measured using a vibrating sample magnetometer (VSM).

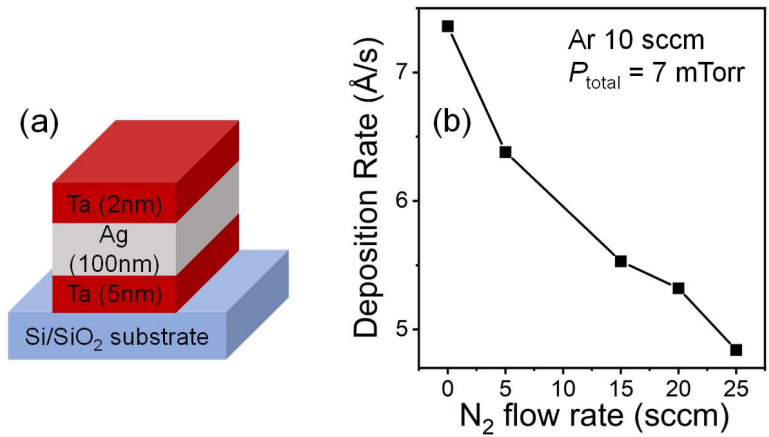

**FIG. 1.** (a) Schematic representation of the sample structure, and (b) $N_2$ flow rate dependence deposition rate of Ag buffer layer.

### III.   Results and Discussion

Figure 1(b) shows the change in the deposition rate of Ag as a function of $N_2$ flow rate. The deposition rate slowed down as the $N_2$ flow rate increased. This is explained by the reduced Ar partial pressure as $N_2$ flow increases under the constant total pressure of 10 mT.

Figures 2(a-e)(i) and (a-e)(ii) show the two-dimensional (2D) XRD profiles of the as-deposited and annealed Ag films deposited in different Ar+$N_2$ atmosphere, respectively. The broadening of the diffraction rings (Debye-Scherrer rings) in the *β* angle direction corresponds

to the distribution of crystallographic orientation of the crystal grains in the out-of-plane direction. The range of the broadening of the *hkl* diffraction peak in the $\beta$ direction is governed by the distribution of the diffraction vector of (*hkl*) plane. Therefore, the rocking curves of the diffraction from the (111) and (002) planes (Figs. 2 (a-e)(iii) and (iv)) and 1D XRD profiles (Figs. 2(f) and (g)), which were obtained by integrating the diffraction intensity to the $\beta$ and $2\theta$ direction, respectively, are also presented. The as-deposited and annealed Ag films grown in the pure-Ar atmosphere (Figs. 2(a)(i) and (ii), respectively) exhibited a highly [111]-oriented bcc structure as indicated by the narrow 111 peak in the $\beta$ direction without the presence of the 002 peak to the substrate normal direction (*i.e.*, $\beta = 180°$). By annealing at 300 °C, the width of the rocking curve of the 111 peak decreased, as shown in Fig. 2(a)(iii), indicating a further enhancement of the [111]-texture. Interestingly, the rocking curve of the 002 peak (Fig. 2 (a)(iv)) showed a splitting into $\beta = 162°$ and $198°$, indicating that there are a small number of grains having the (002) plane to these tilted directions.

By introducing $N_2$ with a flow rate of 5 sccm, a significant change in the crystal orientation of the Ag thin films was observed. The diffraction ring from the (111) plane broadened with noticeable splitting relative to the substrate normal, as clearly seen from the rocking curve shown in Fig. 2(b)(iii). Additionally, a continuous broad diffraction ring from the (001) plane emerged at $2\theta = 44.45°$. The subsequent annealing at 300 °C improved the out-of-plane texture along [001], as shown in Fig. 2(b)(iv). Further increase in the $N_2$ flow rate led to an enhanced [001]-oriented growth of the Ag films, as seen from the increased sharpness of the 002-diffraction ring in the 2D XRD images and the rocking curves from the diffraction of the (002) plane (Fig. 2(b-e)(iv)). The splitting of the 111 peak still existed for all $N_2$ flow rates larger than 5 sccm, which is explained by the formation of twins in the Ag crystals as discussed below based on the STEM observations. However, as shown in Figs. 2(h) and (i), the ratio of the diffraction intensity of the 002 peak to that of the 111 peak ($I_{002}/I_{111}$) and the FWHM of the 002 peak increased and decreased with increasing $N_2$ flow rate, respectively. Therefore, the Ag films obtained an enhanced [001]-texture with lower crystallographic misorientations for higher $N_2$ flow rate. These results demonstrate the crucial role of $N_2$ in promoting the growth of highly [001]-oriented Ag films.

The mechanism of the change in crystal orientation of Ag by the introduction of $N_2$ is not clear at present. As discussed in Ref. [17] and [20], however, a possible mechanism is the influence of $N_2$ on the growth process of Ag thin films during the sputter deposition. The $N_2$ atoms may have acted as a surfactant, promoting a growth of Ag with a specific crystallographic orientation while inhibiting others. The mixture of Ar and $N_2$ in the sputtering gas could influence the surface morphology and the atomic mobility of the Ag atoms, leading to the formation of a texture with a preferred orientation along the [001]-direction.

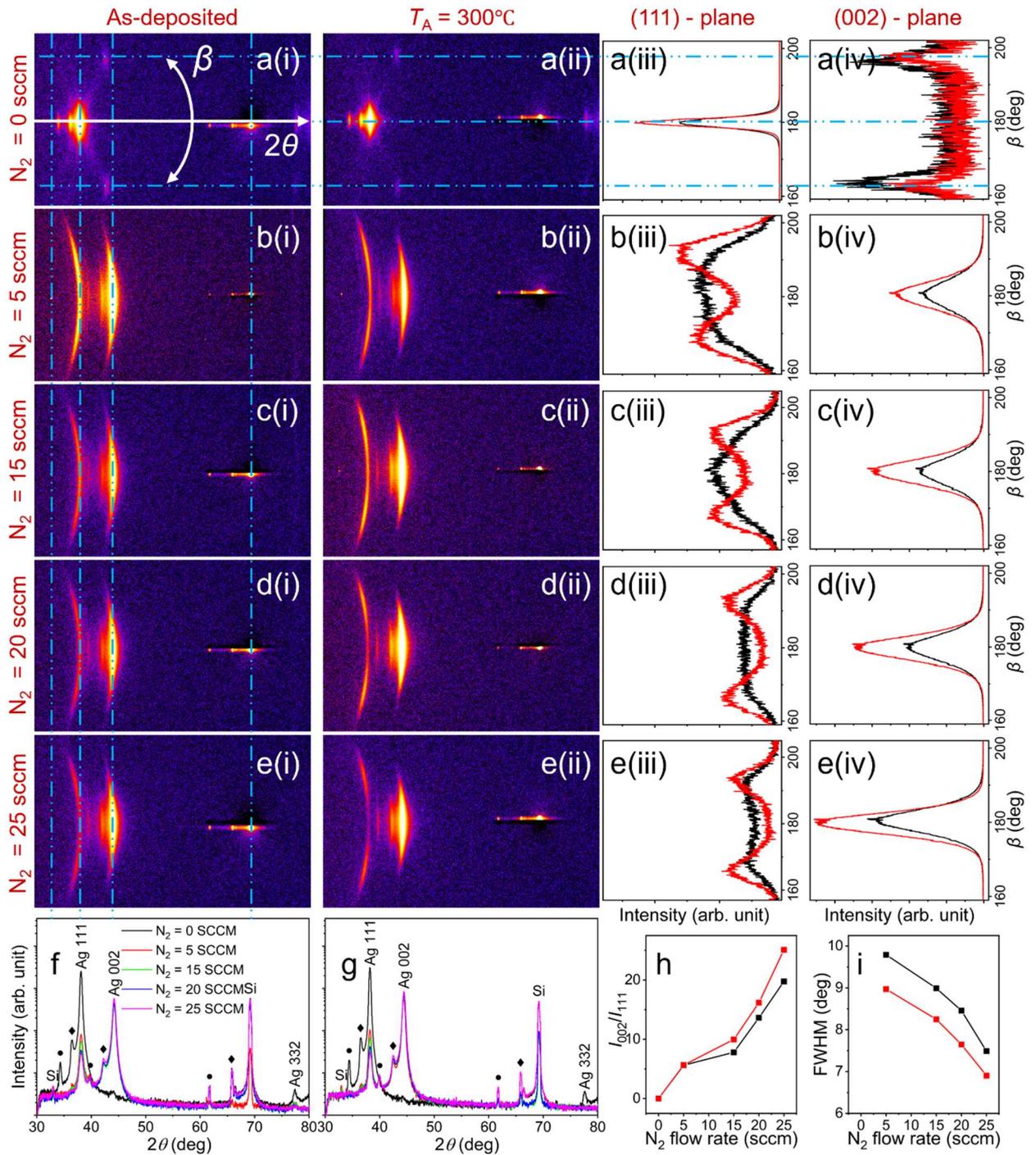

**FIG. 2.** 2D XRD images and corresponding rocking curves from the diffraction along (111) and (002) planes and 1D XRD patterns of Ag thin films grown in different Ar + N$_2$ atmospheres. (a-e)(i) 2D XRD images of as-deposited samples, (a-e)(ii) 2D XRD images of annealed samples, (a-e)(iii) and (a-e)(iv) rocking curve from the diffraction along (111) and (002) planes

respectively. Black and red color graphs represent the as-deposited and annealed sample respectively. (e) 1D XRD pattern converted from the 2D XRD images of the as-deposited samples, (a-e)(i), (f) 1D XRD pattern converted from the 2D XRD images of the annealed samples, (a-e)(ii). The XRD peaks indicated by diamond (♦) and dot (•) symbols correspond to the satellite peaks generated by Cu-$K_{\alpha 1}$ and Cu-$K_\beta$ X-ray wavelength. (g) ratio of the intensities along the [001] and [111]-directions, and (h) FWHM of the rocking curves along the [001]-direction. The $2\theta$ and $\beta$ directions for the 2D XRD plots are represented in Figure a(i).

Next, we studied the effect of the [001]-oriented Ag buffer layer on the crystallographic orientation and structural ordering of a 20-nm-thick CFGG film. Figures 3(a) and (b) show the 2D XRD image of the as-deposited and annealed samples of the layer structure of Si/SiO$_2$/Ta (5 nm)/Ag (100 nm)/CFGG (20 nm)/Ta (2 nm). Figure 3(c) depicts the corresponding 1D XRD pattern. The XRD peaks at $2\theta = 31.31°$ and $65.43°$ correspond to the B2 superlattice 002 and fundamental 004 peaks of CFGG, respectively, indicating that the CFGG film has a [001] out-of-plane orientation with the B2 structure. The lattice constant of the Ag buffer layer and CFGG layer were found to be 0.409 nm and 0.575 nm, respectively. This leads to a small lattice misfit between [001]-oriented Ag and CFGG with an in-plane 45° crystal rotation. The FWHM of the broadening of the 002 diffraction in the $\beta$ direction for the annealed sample shown in Fig. 3(b) was 11.5°, indicating a relatively high [001]-texture of the CFGG film. The inset of Fig. 3(c) displays the results of the magnetization measurement. After annealing, the saturation magnetization, $M_s$ reached ~940 emu/cc. This corresponds to 4.81 $\mu_B$/f.u., which is close to the $M_s$ estimated by the Slater-Pauling rule for this CFGG composition (4.89 $\mu_B$/f.u.).[21]

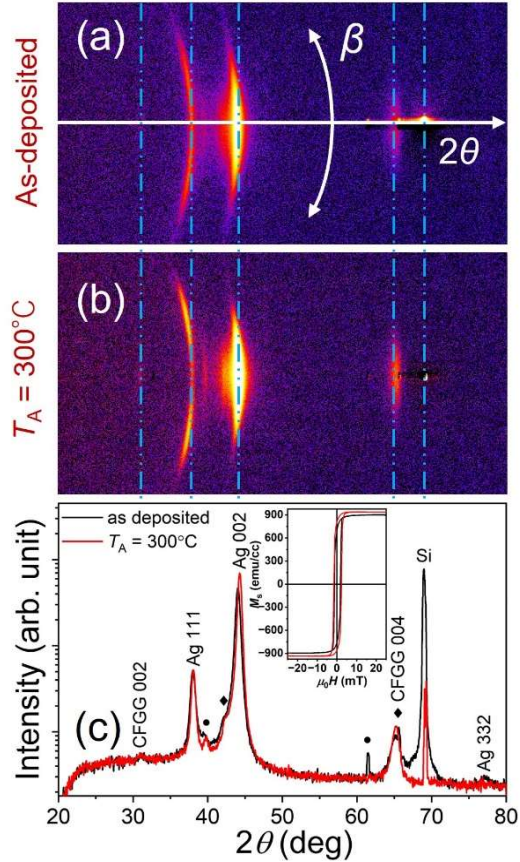

**FIG. 3.** (a), (b) 2D XRD images for the as-deposited and annealed Si/SiO$_2$/Ta (5 nm)/Ag (100 nm)/CFGG (20 nm)/Ta (2 nm) samples, respectively, and (c) the corresponding 1D XRD patterns. The inset in (c) shows the magnetization curves. The XRD peaks indicated by the diamond (♦) and dot (•) symbols correspond to the satellite peaks generated by the Cu-$K_{\alpha 1}$ and Cu-$K_{\beta}$ X-ray radiations. The $2\theta$ and $\beta$ directions for the 2D XRD plots are represented in Figure (a).

Figures 4(a), (b), and (c) display the low-magnification high-angle annular dark-field (HAADF)-STEM image, the corresponding energy dispersive spectroscopy (EDS) elemental map of Ag, Ge, Ta, Au, and Pt, and EDS line profiles for the Si/SiO$_2$/Ta (5 nm)/Ag (100 nm)/CFGG (20 nm)/Ta (2 nm) sample annealed at 300 °C, respectively. Note that Au and Pt were deposited to protect the surface of the TEM specimen during its preparation by FIB/SEM, and EDS line profile was taken across the interfaces, as indicated by an arrow in Fig. 4(b). The HAADF-STEM image showed that the Ag surface was relatively flat, and EDS analysis showed that there was no significant interdiffusion between the constituent layers. The high-magnification HAADF-STEM image in Fig. 4(d) shows that the Ag/CFGG interface is atomically sharp, and the brightly and darkly imaged atomic columns are alternately stacked along the film normal direction throughout CFGG layer, indicating a B2 structure, in which the Fe and (Ga,Ge) atoms are disordered. The B2 structure of CFGG layer is supported by the nanobeam diffraction pattern of CFGG (Fig. 4(e)), which shows the presence of the (002) reflection. From the nanobeam diffraction patterns obtained from the Ag layer, as shown in Fig. 4(e), the B2-ordered CFGG has the epitaxial orientation relationship with the Ag layer described as $(001)_{Ag} \parallel (001)_{CFGG}$ and $[010]_{Ag} \parallel [\bar{1}10]_{CFGG}$, resulting in the [001]-oriented growth of CFGG layer.

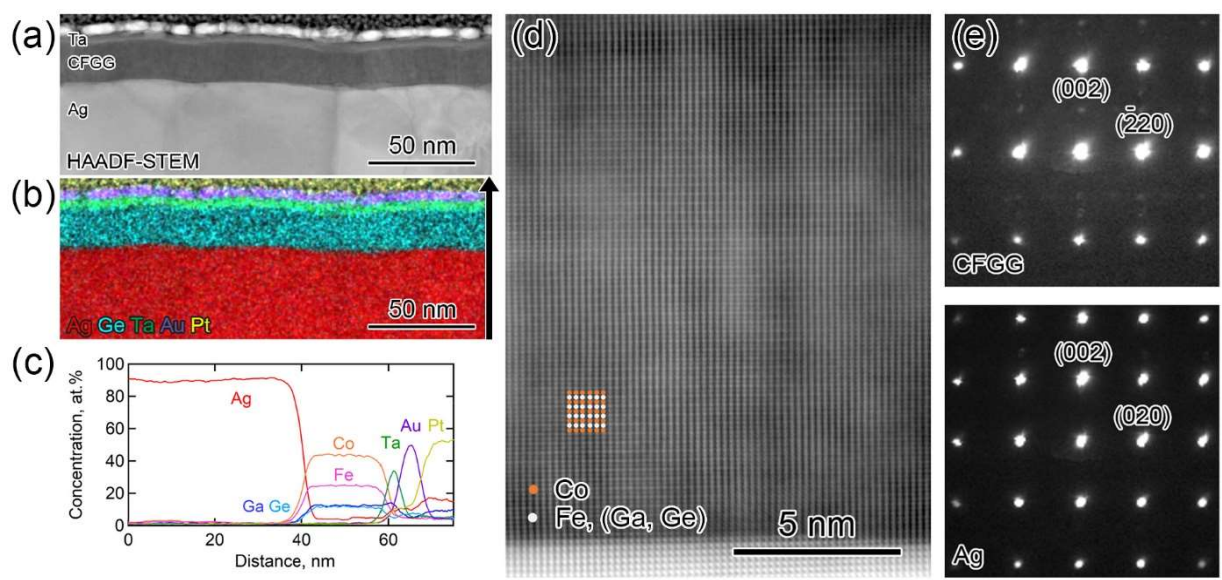

**FIG. 4.** (a) Cross-sectional HAADF-STEM image, (b) EDS map, and (c) EDS line profiles of the Si/SiO$_2$/Ta (5 nm)/Ag (100 nm)/CFGG (20 nm)/Ta (2 nm) sample after annealing at 300 °C. (d) High-magnification HAADF-STEM image of the Ag and CFGG layer taken along the [110] zone axis of CFGG and (e) nano beam diffraction pattern of the CFGG and Ag buffer layer taken from the zone axis of [011] of CFGG and [100] of Ag, respectively.

Next, to investigate the origin of the splitting of the 111 XRD peak of the Ag buffer layer (Figs. 2(b-e)(iii)), microstructural analysis with (S)TEM was performed. Figure 5(a) displays a low-magnification bright field (BF) TEM image of the Si/SiO$_2$/Ta (5 nm)/Ag (100 nm)/CFGG (20 nm)/Ta (2 nm) sample annealed at 300 °C. The difference in diffraction contrasts reveals the polycrystalline nature of the Ag buffer layer, which has well-defined interfaces between adjacent layers and a uniform thickness. As shown in Fig. 5(b), the nanobeam diffraction pattern taken from the large area of the Ag buffer layer surrounded by the solid line in Fig. 5(a) clearly shows the [001]-oriented growth of Ag. There is also a stripe-like diffraction contrast tilted ~59.4° from the film plane within a Ag grain, as indicated by the arrows in Fig. 5(a). As seen in the nanobeam diffraction pattern taken from the area surrounded by the dashed line, Fig. 5(c), and the key diagram, Fig. 5(d), the two rhombohedral constructions in the white/black and red dashed lines are mirrored with respect to the ($\bar{1}$11) plane, indicating the twin typically observed in Ag. In addition, since the ($\bar{1}$11) plane makes an angle of 54.74° with the (002) plane, *i.e.*, the film plane, the stripe-like diffraction contrast in Fig. 5(a) is attributed to the (1$\bar{1}$1) trace of the twin boundaries. The key diagram also shows that the ($\bar{1}$1$\bar{1}$) plane of the twinned region is tilted 15.78° from the (002) plane of the parent grain. The high magnification HAADF-STEM in Fig. 5(e), including the parent grain and the twinned region, shows that the ($\bar{1}$1$\bar{1}$) plane of the twinned region is tilted ~15.0° from the

(002) plane of the parent grain. Therefore, the splitting of the 111 peaks in the XRD profile is attributed to the (111) plane of the twinned region. No twin was observed in the CFGG layer.

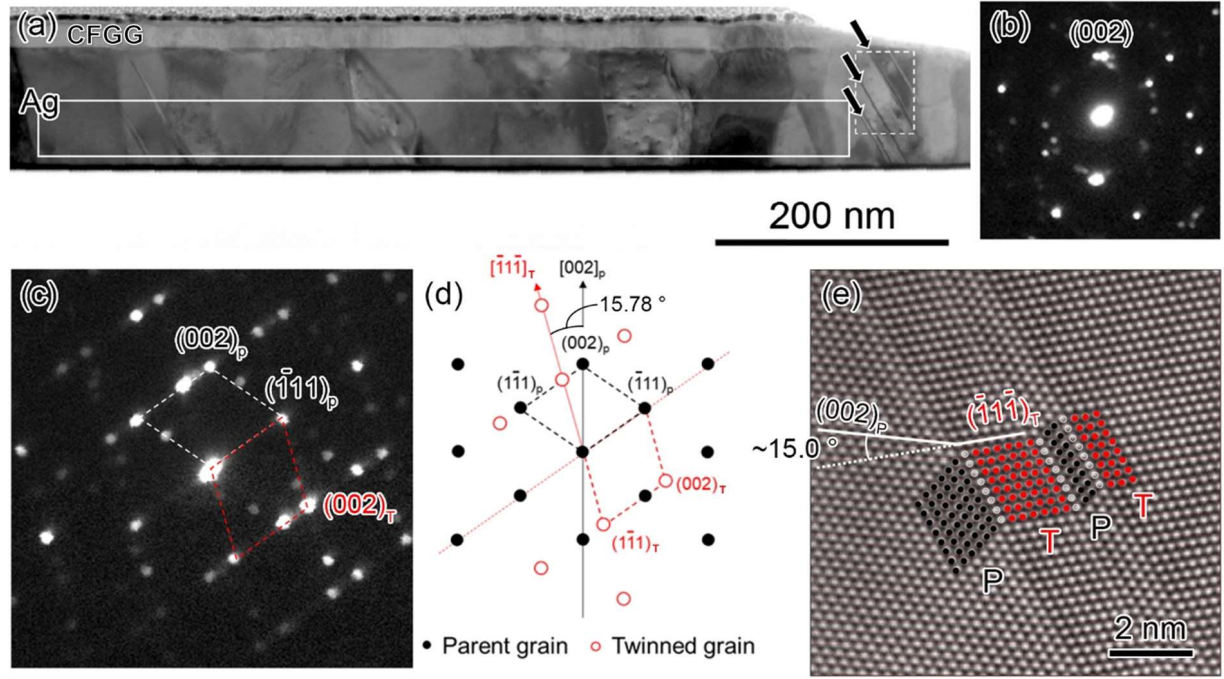

**FIG. 5.** TEM analysis of Si/SiO$_2$/Ta (5 nm)/Ag (100 nm)/CFGG (20nm)/Ta (2 nm) after annealing at 300 °C. The Ag buffer layer was grown at N$_2$ flow rate of 25 sccm. (a) low-magnification bright field (BF) TEM image, nano beam diffraction pattern of the silver buffer layer obtained from the area surrounded by (b) solid and (c) dashed lines, (d) key diagram that helps interprets the nanobeam diffraction pattern, (c), and (e) atomic-resolution HAADF-STEM image including parent grain and twinned region. Note that "P" and "T" in (c) through (e) stand for parent and twin, respectively.

## V. Conclusions

This study demonstrates a simple method to control the preferred crystal orientation of Ag and CFGG Heusler alloys films. By introducing $N_2$ during the sputter deposition of Ag, we obtain a highly [001]-oriented Ag buffer layer with a relatively flat surface. This allows the growth of a B2-ordered CFGG Heusler alloy thin film with a [001] out-of-plane texture. This technique provides the use of [001]-oriented Heusler alloy films for spintronic devices for industrial applications. This approach may also be extended to change the crystallographic orientation of various other fcc metals, particularly those with low reactivity towards $N_2$ (e.g., Au, Pt, Cu, Ni) by fine tuning the deposition condition.

**Acknowledgments**

The authors thank Sanae Kuramochi for her technical assistance in making the samples and the initial characterization. This work was partly supported by a collaborative research project with TDK Corporation, JSPS KAKENHI Grant No. 21H01608, and MEXT Initiative to Establish Next-generation Novel Integrated Circuits Centers (X-NICS) Grant No. JPJ011438.

**Author Declarations**

The authors have no conflicts to disclose.

**Data Availability Statement**

The data supporting the findings of this study are available from the corresponding author upon reasonable request.